\documentclass[onecolumn,prc,showpacs,amsmath,amssymb,epsfig,10pt]{revtex4}
\usepackage{epsf,epsfig,graphicx}
\begin{document}
\titlepage
\title{ The Observable of Lambda Polarity Caused by Source Global Angular Momentum Localizing in Peripheral Au-Au Collision at RHIC }
\author{ X. Sun$^a$, Z. Yang$^{b,c}$\\
$^a$Institute of High Energy Physics, Beijing 100039, China\\
$^a$Graduate University of the Chinese Academy of Sciences, Beijing 100039, China\\
$^b$Department of Engineering Physics, Tsinghua University,
Beijing 100084, China\\
$^c$Center of High Energy Physics, Tsinghua University, Beijing
100084, China}

 \begin{abstract}
An observable to  measure the polarity of lambda that caused by
source global angular momentum in peripheral AU-AU collision at RHIC
is proposed.This observable's capacity of measurement is tested by
Monte Carlo method.And the main factors that influence the
observable are also researched.This observable will give an
effective proof of the formation of deconfined matter.
\end{abstract}
 \maketitle
\newpage

\section{Introduction}
As we all know,only a few of events are center-to-center in Au-Au
collision in RHIC because the nucleous has enough volume to be
regarded as a ball rather than a point.Most of events are peripheral
that is only a part of the nuclei collide to each other when two
nuclei meet.Just as two macroscopical bodies collide peripherally to
each other,a global angular momentum will be reserved in the
overlapped region in heavy ion collision.It was firstly proposed by
Liang and Wang\cite{wang:2004xn,Liang:2004xn}. At the same time the
unoverlapped parts keep on moving in initial direction and the
overlapped part i.e. the source becomes still for the loss of
momentum.The questions in front of us is how to verify the existence
of the  global angular momentum and how to measure  it in experiment
if it exists.

Unfortunately,the first question can not be replied directly because
all information about the source we can get is from the final-state
particles. On the other hand,many results from both experiment and
theory show that an deconfined matter created after the collision.
The first evidence comes from "jet quenching"\cite{PHEN02,STAR02}.It
was predicted to occur as a result of energy loss by the hard
scattered partons due to interactions with the surrounding dense
medium \cite{Bjo82,Tho91,Gyu94}.The theory of this energy loss has
been a topic of intense research over the past few years
\cite{Zak96,BDMPS,Wie00,BDMS,BSZ00}\cite{berndt}.The second evidence
is the elliptic flow found in heavy ion cillision. Quark coalescence
model\cite{alcormicor}\cite{Lin:2001zk} has been applied to describe
the elliptic flow at RHIC for different flavors
\cite{Lin:2002rw,Voloshin:2002wa,Molnar:2003ff}\cite{Molnar}.Another
surprising early measurements from RHIC shows that the proton/pion
ratio reaches or even exceeds unity for transverse momenta $p_t$
above 2 GeV/c \cite{poverpidata}. One explanation for this phenomena
is that quarks originating from different nucleon-nucleon collisions
recombine via coalescence mechanisms
\cite{voloshin,muellergroup,friesasymmetry,kogroup,molnar}\cite{Subrata}.If
an matter composed by partons created in heavy ion collision,The
global angular momentum between partons can convert to spin of
final-state particles during the formation of these
particles\cite{wang:2004xn,Liang:2004xn}.The reason is that the
orbital angular momentum of partons is part of spin of particle .So
if the source has global angular momentum ,this angular momentum
will localize to final-state particles and the particles will be
polarized.But if the matter is still confined,the polarization will
not happens.

 As experimental result supports the formation of deconfined
matter,we go to the second question. Now the second question
converts to how to measure
 the polarity of some kind of particle.This question will mainly be
 discussed in this paper:1,choosing particle and
 observable;2,arguing  the relation between the observable and
 polarity;3, factors influencing the measurement.

\section{choosing particle and observable}
There are some criterions in choosing the particle to reflect the
global angular momentum of the source.Firstly,its spin can't be zero
and its decay-length must be suitable to reconstruct. Secondly,it
must decay by weak interaction and its daughter's angular
distribution must be spherical asymmetrical because of the break-out
of the conservation of parity in weak interaction. Finally,there
must be many such particles produced in heavy ion collision so the
mass of the particle should not too large. Under these
criterions,$\Lambda/\bar{\Lambda}$ is a good
candidate\cite{wang:2004xn,Liang:2004xn}.

After the particle is chosen the observable of the particle should
be designed subsequently.We define a reaction plane in heavy ion
collision by the beam direction and the  impact parameter vector and
we can fix it in experiment, but we can't distinguish which
direction is up or down for the polarized particles if they are
indeed polarized by the global angular momentum of the source. The
distribution of the daughter  of $\Lambda$ (i.e. $\pi^{-}$ and
proton) is written as

\begin{eqnarray}
 \label{daughter distribution}
P(\theta)=\frac{dN}{d\cos\theta}=\frac{1}{2}(1+\alpha
P_{\Lambda}\cos\theta)
\end{eqnarray}

$\theta$ is the polar angle in the  center of mass frame  of
polarized  $\Lambda$ when we define the spin direction of $\Lambda$
as the z axis. $\alpha $ is a constant for $\Lambda$ decay and it
equals 0.642.$P_{\Lambda}$ is the polarizability of $\Lambda$. If
all the event after turned to reaction plane  are counted we will
still get a symmetrical distribution of daughters of $\Lambda$ in
the center of  mass frame of $\Lambda$ because the second term of
right of (\ref{daughter distribution}) will give a negative sign if
the change of polar angle $\theta$ is $\pi$.To overcome this
difficult,we define observable as following.

Firstly we should fix the reaction plane in experiment for an
event.The reaction plane divide the x-y plane into two parts.We call
one of them up direction and the other down direction.The reaction
is immovable for the event and the $\Lambda$ in it.For the event of
$N \Lambda s$ we can count the number of $\Lambda$ s whose one of
the daughters such as proton belongs to the up or down direction in
the center  of mass  frame  of $\Lambda$ as $N_{\uparrow}$ and
$N_\downarrow$.It is easy to know $N=N_{\uparrow}+N_\downarrow$.We
define $M_{N}(i)$ for the i-th event of $N \Lambda$ s as
\begin{eqnarray}
 \label{event minus}
M_{N}(i)=\mid N_{\uparrow}-N_\downarrow\mid
\end{eqnarray}
.Then we add $M_{N}(i)$ for all the events of $N \Lambda s$together
to get $M_{N}$ .$N_{event}$ is the number of all events of $N
\Lambda s$.We can get
\begin{eqnarray}
 \label{observable}
P(N)=\frac{M_{N}}{N_{event}}
\end{eqnarray}
The $P(N)$ is the observable to reflect the global angular momentum
for the event whose number of $\Lambda$ is $N$.

The next question asked instantly is why and how $P(N)$  reflects
the existence of the global angular momentum of the source (i.e.the
polarity of $\Lambda$ emitting from it ).Qualitatively, if $\Lambda$
emitting from the source is polarized and the reaction plane has
been confirmed, the numbers of $\Lambda$ that crosses two sides of
the reaction plane are unequal because of (\ref{daughter
distribution}).The difference between  them is $M_{N}(i)$  for the
i-th event of $N \Lambda$ s. So it is obvious the relation between
$P(N)$ and the polarization because of  the relation between $P(N)$
and $M_{N}(i)$.Concretely ,we  calculate the value of $P(N)$ for the
unpolarized event firstly,then contrast the value from real data to
it.

\section{the relation between the observable and
 polarity}
To explain this issue,the first value should be calculate is the
$P(N)$ for the unpolarized event.In unpolarized event,the spin
direction of $\Lambda$ is random.So it is easy to get:
\begin{eqnarray}
 \label{unpolarized}
P(N)=\frac{\sum_{m=0}^{m=n}\mid 2m-n\mid C_n^m}{\sum_{m=0}^{m=n}
C_n^m}
\end{eqnarray}
The value of $P(N)$ is listed below:
\begin{figure}[ht]
\vspace*{+0cm} \centerline{\epsfxsize=6cm \epsffile{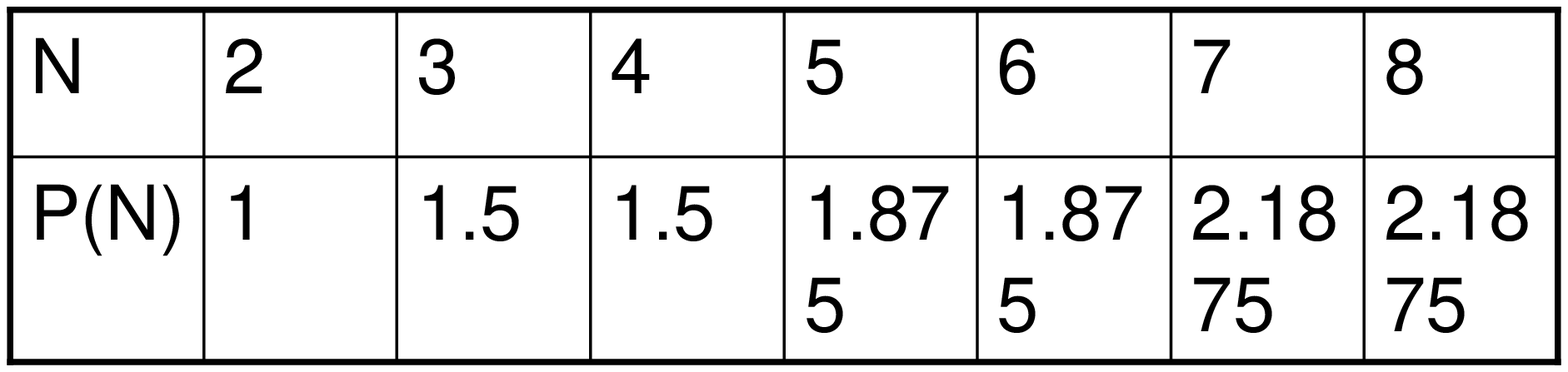}}
\caption{\label{Fig.1} The value of P(N)in unpolarized event }
\end{figure}
(\ref{unpolarized}) can be understand in this way:Let's just see
proton decayed from $\Lambda$.  The probability of there are $m$ $
\Lambda$ whose proton belong to up direction of the reaction plane
is $\frac{C_n^m}{\sum_{m=0}^{m=n}C_n^m}$.$\mid
N_{\uparrow}-N_\downarrow \mid$ is $\mid m-(n-m)\mid$ i.e.$\mid
2m-n\mid$ for event of $N$ $\Lambda$.Add the probability multiplied
by $\mid 2m-n\mid$ for $m$ from 0 to $N$ and the expression of
$P(N)$ can be got as (\ref{unpolarized}).

Calculating the value of $P(N)$ in unpolarized event,we should
estimate the $P(N)$ and the variance of it in polarized event for
the statistical quantity that we can get from experiment.We can do
that just like we define $P(N)$ by use of Monte Carlo.After
selecting the events that contain $N$ $\Lambda$s and turning to the
center of mass frame
 of each $\Lambda$,we produce direction of proton from
$\Lambda$ satisfying the distribution of (\ref{daughter
distribution}) regarding x axis  as the reaction plane for all
events.Then count the number of protons that lies in the first and
second quadrants as $N_\uparrow$ and that in the third and fourth
quadrants as $N_\downarrow$.After counting and  adding the
difference of these two value (i.e. $M_N(i)$) for all event of $N$
$\Lambda$s, we can get $P(N)$ by use of (\ref{observable}).This
process should be repeated for many times for example 50 times to
get the variance of $P(N)$.The pictures below show the result

\begin{figure}[ht]
\vspace*{+0cm} \centerline{\epsfxsize=6cm \epsffile{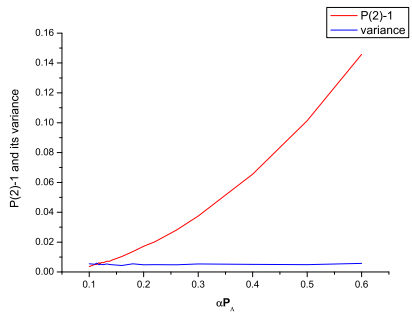}}
\caption{\label{Fig.2}Monte Carlo result for N=2,the red line is
P(2)-1,the blue line is the variance of P(2) for 50 times }
\end{figure}
\begin{figure}[ht]
\vspace*{+0cm} \centerline{\epsfxsize=6cm \epsffile{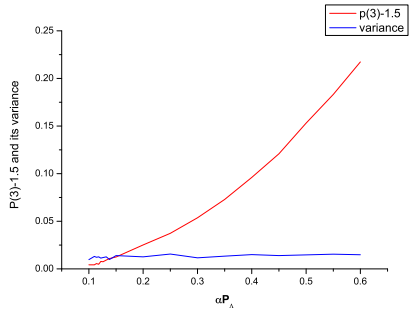}}
\caption{\label{Fig.3}Monte Carlo result for N=3,the red line is
P(3)-1.5,the blue line is the variance of P(3) for 50 times }
\end{figure}

The accuracy of this method is determined by the ratio of
$P(N)-P_r(N)$ to the variance of $P(N)$. $P_r(N)$ means the
reference value for $P(N)$ in unpolarized events listed in Fig.1.
Presumedly if the degree of confidence is set to 3$\sigma$,we can
get signal in case of $\alpha P_\Lambda >$0.2.This estimate is from
Fig.2 and Fig.3 with no other factors considered.

\section{factors influencing the measurement}
The value of $P(N)$ is affected by a lot of factors.We will discuss
the main ones in this section.Obviously,the resolution of reaction
plane can influence $P(N)$ severely.If the real reaction plane is
known,the degree of asymmetry between up and down the reaction plane
is the maximum of whatever resolution of reaction plane.On the other
hand,if the resolution of reaction plane is $\pi$,in other words,we
don't know the reaction plane,The value of $P(N)$ will equal that in
unpolarized events.Picture below gives the trend of $P(2)$ at
$\alpha P_\Lambda$=0.2.

\begin{figure}[ht]
\vspace*{+0cm} \centerline{\epsfxsize=6cm \epsffile{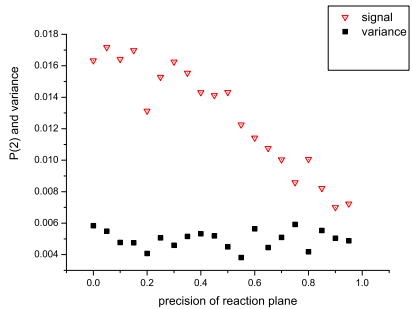}}
\caption{\label{Fig.4}Monte Carlo result for N=2,the red triangle is
P(2)-1,the black square is the variance of it for 50 times,X-axis
means the reaction plane resolution divided by $\pi$ }
\end{figure}
From Fig.4 we can know that if this method is feasible at $N$ =2 and
$\alpha P_\Lambda$=0.2  the resolution of reaction plane must be
smaller than 0.4$\pi$. This is a quite loose requirement.

Another factor is the resolution of momentum.The direction of proton
decayed from $\Lambda$ is measured in the  center of mass frame and
the transformation from laboratory frame to the  center of mass
frame needs the momentum of both proton and pion decayed from
$\Lambda$.So the resolution of momentum will affect the direction of
daughters of $\Lambda$.the picture below shows the trend of $P(2)$
depending on the  momentum  resolution at $\alpha P_\Lambda$=0.2 .
\begin{figure}[ht]
\vspace*{+0cm} \centerline{\epsfxsize=6cm \epsffile{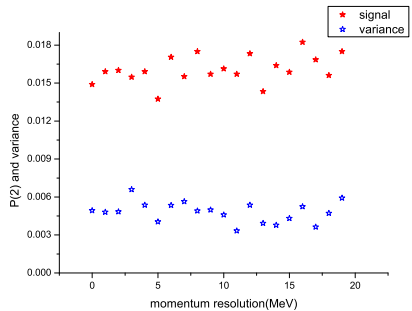}}
\caption{\label{Fig.5}Monte Carlo result for N=2,the red pentacle is
P(2)-1,the blue empty pentacle is the variance of it for 50
times,X-axis means the momentum resolution   }
\end{figure}
From Fig.5 we can know that the variety of $P(2)$ depending on the
momentum resolution from 0 to 20 MeV is unconspicuous. So the
influence from momentum resolution should be small.

There are also other factors that affect $P(N)$ ,such as the purity
of $\Lambda$.We can estimate the influence of it from the
reconstruction of $\Lambda$.

\section{Conclusions}
An observable $P(N)$ to measure the source global angular momentum
localization  in peripheral AU-AU collision at RHIC is proposed in
this article.And the relation between $P(N)$ and the polarizability
is shown by simulation in which the quantity of statistics is from
real data of AU-AU collision in 200 GeV at STAR in 2001.The variance
of $P(N)$ is also shown by this.Then the factors that can influence
$P(N)$ are discussed.The influence of momentum resolution is trivial
within the resolution of STAR detector . From simulation result, the
resolution of reaction plane should be smaller than 0.4$\pi$ for a
feasible measurement.If $\alpha P_{\Lambda}$ is bigger than 0.2
i.e.$P_{\Lambda}>$0.312,this method can give a meaningful result
with the degree of confidence about 3$\sigma$.If $\alpha
P_{\Lambda}$ is smaller than 0.2,this method can give an upper limit
of $P_{\Lambda}$.

{\bf Acknowledgments:} We thank  Dr J. Fu for supporting us in many
respects and their constant helps. We thank Prof. J. Li and Z. Zhang
for the earnest supervision. We thank Prof.Shaomin Chen and Yuanning
Gao for the fruitful discussions in this work. The work was
supported in part by the grants NSFC 10447123.


\end{document}